\documentclass[11pt,a4paper]{article}
\usepackage{jheppub_kim}

\usepackage{pdflscape}
\usepackage{amsmath}
\usepackage{amssymb}
\usepackage{dcolumn}
\usepackage{bm}
\usepackage{color}
\usepackage{epsfig}
\usepackage{amsfonts}
\usepackage{graphicx}
\usepackage{subfigure}
\usepackage{dcolumn}

\newcommand{\be}{\begin{equation}}
\newcommand{\ee}{\end{equation}}
\newcommand{\bea}{\begin{eqnarray}}
\newcommand{\eea}{\end{eqnarray}}

\def\be{\begin{equation}}
\def\ee{\end{equation}}
\def\bea{\begin{eqnarray}}
\def\eea{\end{eqnarray}}

\begin{document}

\title{Logarithmic corrected $F(R)$ gravity in the light of Planck 2015}

\author[a]{J. Sadeghi}
\author[a]{H. Farahani}

\affiliation[a]{Department of Physics, University of Mazandaran, P .O .Box 47416-95447, Babolsar, Iran}

\emailAdd{pouriya@ipm.ir, h.farahani@umz.ac.ir}

\abstract{In this letter, we consider the theory of $F(R)$ gravity with the lagrangian density $ \pounds = R+\alpha R^2 + \beta R^2 \ln \beta R $. We obtain the constant curvature solutions and find the scalar potential of the gravitational field. We also obtain the mass squared of a scalaron in the Einstein$^,$s frame. We find cosmological parameters corresponding to the recent Plank 2015 results. Finally, we analyze the critical points and stability of the new modified theory of gravity and find that logarithmic correction is necessary to have successful model..
}


\maketitle

\section{Introduction}
Recent astrophysical observations clarify the accelerated expansion of universe \cite{Riess:1998cb,Riess:2004a,Perlmutter:1998np}, which may be described by dark energy scenario. In that case, there are several dark energy models, the simplest one is the cosmological constant, however it is not a dynamical model, so there are another alternative theories such as quintessence  \cite{Ratra:1987rm,Wetterich:1987fm,Liddle:1998xm,Guo:2006ab,Khurshudyan:2014a,Dutta:2009yb}, phantom \cite{Caldwell:1999ew,Caldwell:2003vq,Nojiri:2003vn,Onemli:2004mb,
Saridakis:2008fy,Saridakis:2009pj,Gupta:2009kk}, and quintom \cite{Guo:2004fq,Zhao:2006mp,Cai:2009zp} models, or holographic dark energy proposal \cite{Li:2010a,Sadeghi:2014a,Setare:2007azp,Saridakis:2008azp}. Moreover, there are interesting models to describe the dark energy such as Chaplygin gas \cite{P89,P92,P93,P96,P97,P100-1,P100-2,P102,P103,P1002,P1003,PLB636(2006)86,0812.0618,P26 1103.4842,IJTP52(2013)4583,PLB646(2007)215,P59 1012.5532,P60 1102.1632,ASS341(2012)689}.\\
Modification of the Einstein-Hilbert (EH) action through the Ricci scalar can describe inflation and also present accelerated expansion of universe. This called $F(R)$ gravity model, so there are several ways to construct a $F(R)$ gravity models \cite{0612180,0912.5474,1402.2592,1203.5220}. In this paper we consider the particular case of the $F(R)$ gravity model where the Ricci scalar replaced by a new function,
\begin{equation}\label{I1}
F(R) = R+\alpha R^2 + \beta R^2 \ln \beta R,
\end{equation}
where $ \beta>0 $ is the parameter with the squared length dimension and also $\alpha>0$. This model can describe the universe evolution without
introducing the dark energy \cite{1011.0544}, where the cosmic acceleration exists due to the modified gravity. So, $F(R)$ gravity
models can be replaced to the cosmological constant model.\\
The function given by (\ref{I1}) without logarithmic correction ($\beta=0$) has been studied by \cite{1003.3179,1301.5189} which is applicable to a neutron star with a strong magnetic field \cite{1304.1871}. In order to consider effect of gluons in curved space-time, the logarithmic correction in (\ref{I1}) proposed by \cite{1307.7977}. In the Ref. \cite{1307.7977} a phenomenological model based on the equation (\ref{I1}) proposed. Motivated by this model, we would like to use relation (\ref{I1}) to study some cosmological parameters in the light of new data of Planck 2015.\\

Initial idea of the $F(R)$ gravity models successfully examined by Refs. \cite{0705.1158,0705.3199,0706.2041}. Then, several models of $F(R)$ gravity introduced in the literatures \cite{1202.4807,1204.6709,1310.6915,0712.4017,0905.2962,1006.1769,1106.3569}. These are indeed phenomenological models which describe evolution of universe. The Minkowski metric $ \eta_{\mu\nu}=diag(-1, 1, 1, 1)$ and $c = \hbar =1$ are used in the initial $F(R)$ gravity model \cite{PLB91}. Now, we would like to use logarithmic corrected $F(R)$ model given by the equation (\ref{I1}) and exam cosmological consequences of the model using recent data of Planck \cite{2015}.\\

This paper is organized as follows. In section 2, we introduce the model, then study constant curvature condition in section 3. In section 4 we obtain form of the scalar tensor. Cosmological parameters like tensor to scalar ratio obtained in section 5. Critical points and stability analyzed in section 6. Finally, in section 7 we give conclusion.

\section{The Model}
We begin with the equation (\ref{I1}) to modify the Ricci scalar $R$ in the EH action. The
function $F(R)$ satisfies the conditions $F(0)= 0,$ corresponding to the flat space-time without cosmological constant.  Thus, the
action in the Jordan frame becomes,
\begin{equation}\label{M1}
S = \int d^4 x \sqrt{-g} \pounds  = \int d^4 x \sqrt{-g}
\left[\frac{1}{2\kappa^2} F(R)+ \pounds_{m} \right ],
\end{equation}
where $\kappa = M^{-1}_{pl}$, and $M_{pl}$ is the reduced Planck mass, and $\pounds_m$ is the matter Lagrangian density. Our main goal is to study the cosmological parameters
describing inflation and the evolution of the early universe. However, we can discuss about consequences in the late time. From the equation (\ref{I1}) we obtain,
\begin{eqnarray}\label{M2}
F^{\prime}(R) &=& 1 + \gamma R + 2 \beta R \ln \beta R, \nonumber\\
F^{\prime \prime}(R) &=& \lambda + 2 \beta \ln \beta R,
\end{eqnarray}
where $\gamma = 2\alpha + \beta $ and $\lambda = 2\alpha+3\beta$.
The function $F(R)$ obeys the quantum stability condition $F^{\prime \prime}(R) > 0$ for $ \alpha > 0$ and $ \beta >
0$. This ensures the stability of the solution at high curvature.
It follows from the equation (\ref{M2}) that the condition of classical stability
$F^{\prime}(R) > 0$ leads to,
\begin{equation}\label{M3}
1 + \left(\gamma + 2 \beta \ln \beta R \right) R > 0,
\end{equation}

\section{Constant curvature condition}
We consider constant curvature solutions of the equations of motion
that follow from the action given by (\ref{M1}) without matter. The governing equation is
given by \cite{B},
\begin{equation}\label{CCC1}
2 F(R) - R F^{\prime}(R) = 0,
\end{equation}
and hence,
\begin{equation}\label{CCC2}
R = \frac{1}{\beta},
\end{equation}
which satisfy $ 0 < \beta R < 1. $ Here, the condition
$\frac{F^{\prime}(R)}{F^{\prime \prime}(R)}
> R $, is satisfied, and therefore the model can describe primordial and present dark
energy, which are future stable. From the equation (\ref{M2}), we obtain,
\begin{equation}\label{CCC3}
\frac{F^{\prime}(R)}{F^{\prime \prime}(R)}=\frac{1 + \gamma R + 2 \beta R \ln \beta R}{\lambda + 2 \beta \ln \beta
R} > R,
\end{equation}
which simplifies to $ \beta R < \frac{1}{2}$. Thus, the solution
$R _{0}= 0$ satisfy the equation (\ref{I1}) which then imply that the flat space-time is stable. The
second constant curvature  solution $ \beta R_{0} \approx 1$ does
not satisfy the equation (\ref{CCC3}), and this leads to unstable de Sitter space-time, so describes inflation.

\section{The scalar tensor form}
\begin{figure}[h!]
 \begin{center}$
 \begin{array}{cccc}
\includegraphics[width=60 mm]{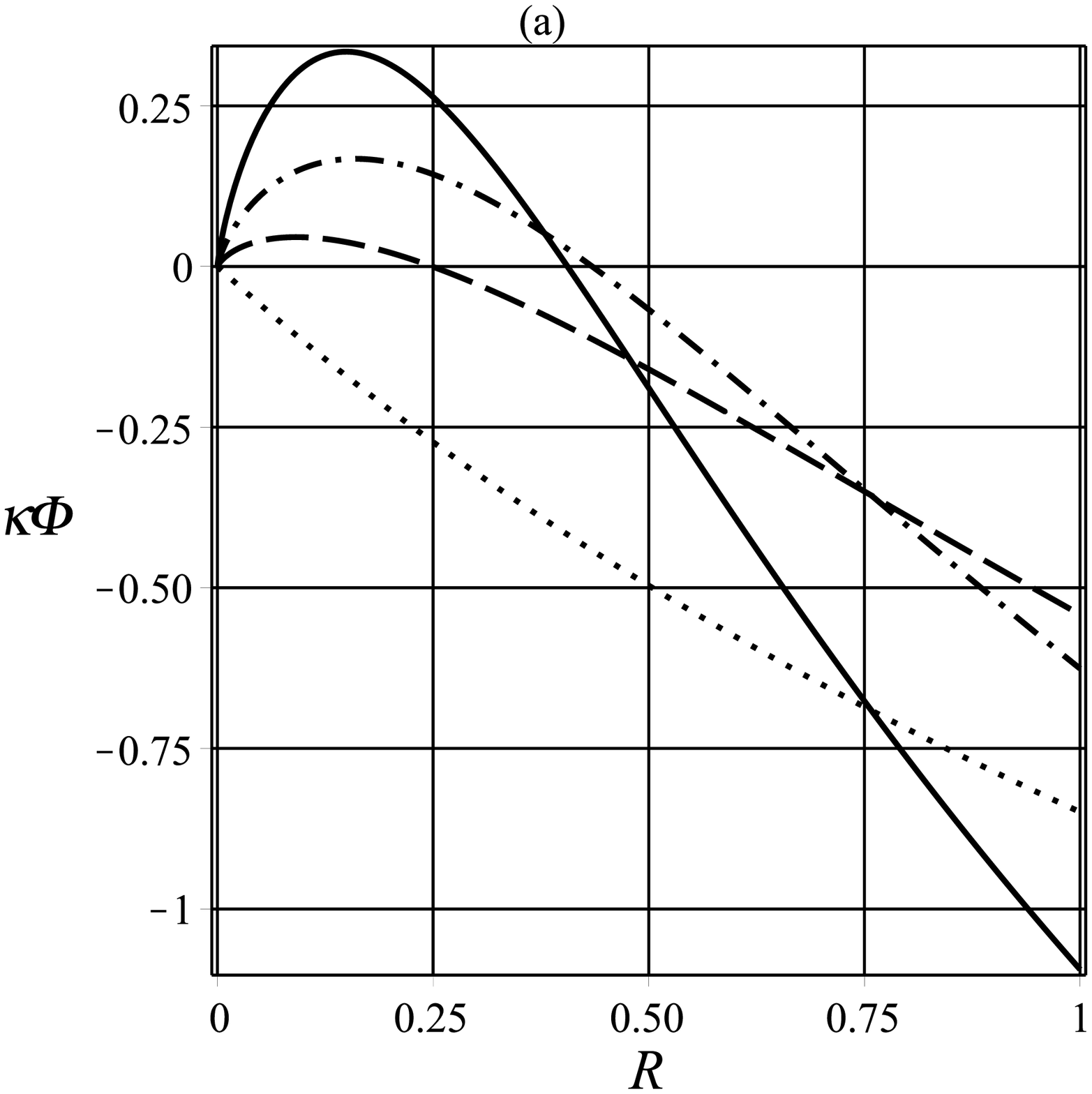}\includegraphics[width=60 mm]{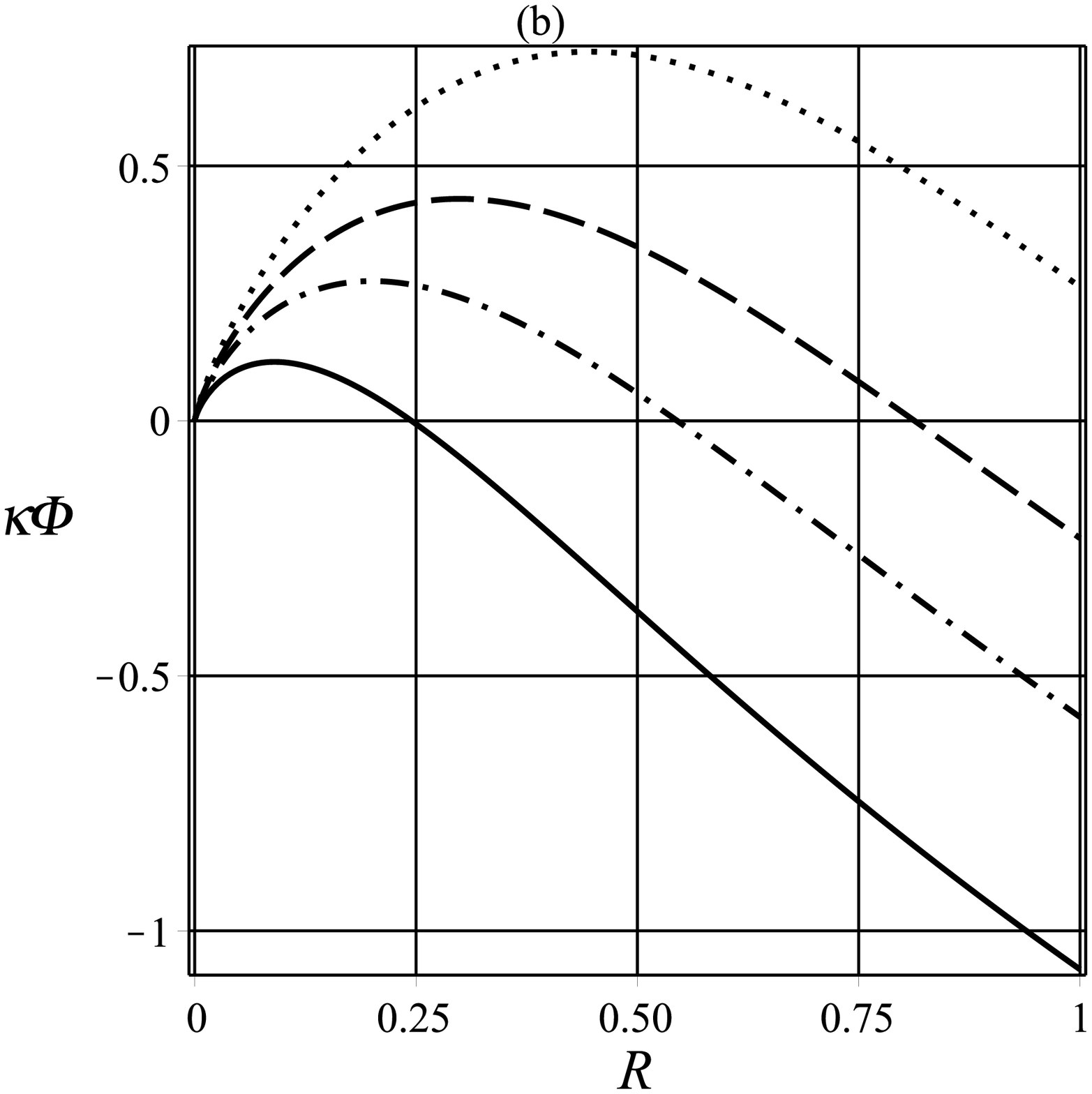}
 \end{array}$
 \end{center}
\caption{The function $\kappa \Phi$ versus $R$. (a) $\alpha=0.5$, $\beta=0$ (dot), $\beta=0.2$ (dash), $\beta=0.4$ (dot dash), $\beta=0.8$ (solid). (b) $\beta=0.5$, $\alpha=0$ (dot), $\alpha=0.2$ (dash), $\alpha=0.4$ (dot dash), $\alpha=0.8$ (solid).}
 \label{fig1}
\end{figure}

In the Einstein frame corresponding to the scalar tensor theory of
gravity, we have the following conformal transformation of the metric \cite{9312008},
\begin{equation}\label{S1}
\tilde{g}_{\mu \nu} = F^{\prime}(R)g_{\mu \nu} = (1 + \gamma R + 2
\beta R \ln \beta R ) g_{\mu \nu}.
\end{equation}
In that case the action given by the equation (\ref{M1}) with $\pounds_m = 0$ written as,
\begin{equation}\label{S2}
S =  \int d^4 x \sqrt{-g} \left[\frac{1}{2\kappa^2} \tilde{R} -
\frac{1}{2} \tilde{g}^{\mu \nu} \nabla_{\mu} \Phi \nabla_{\nu}\Phi -
V(\phi) \right ],
\end{equation}
where $\nabla_{\mu}$ is the covariant derivative, and $\tilde{R}$ is determined using the conformal metric in the equation (\ref{S1}). The scalar field $\Phi$
was found to be,
\begin{equation}\label{S3}
\Phi = -\sqrt{\frac{3}{2}} \frac{\ln({1 + \gamma R + 2 \beta R \ln
\beta R})}{\kappa}.
\end{equation}
In Fig. \ref{fig1} we plotted the function $\kappa \Phi(R)$ for different values
of $ \beta $ and $ \alpha$. From Fig. \ref{fig1} (b) we can see that increasing $\alpha$ decreases value of the scalar field, while there is no regular behavior with variation of $\beta$. However the condition $0 < \beta R < 1 $ satisfied in the plots. We can see that the scalar field (multiple by $\kappa$) is positive for small $R$ and $\beta>0$.\\

\begin{figure}[h!]
 \begin{center}$
 \begin{array}{cccc}
\includegraphics[width=60 mm]{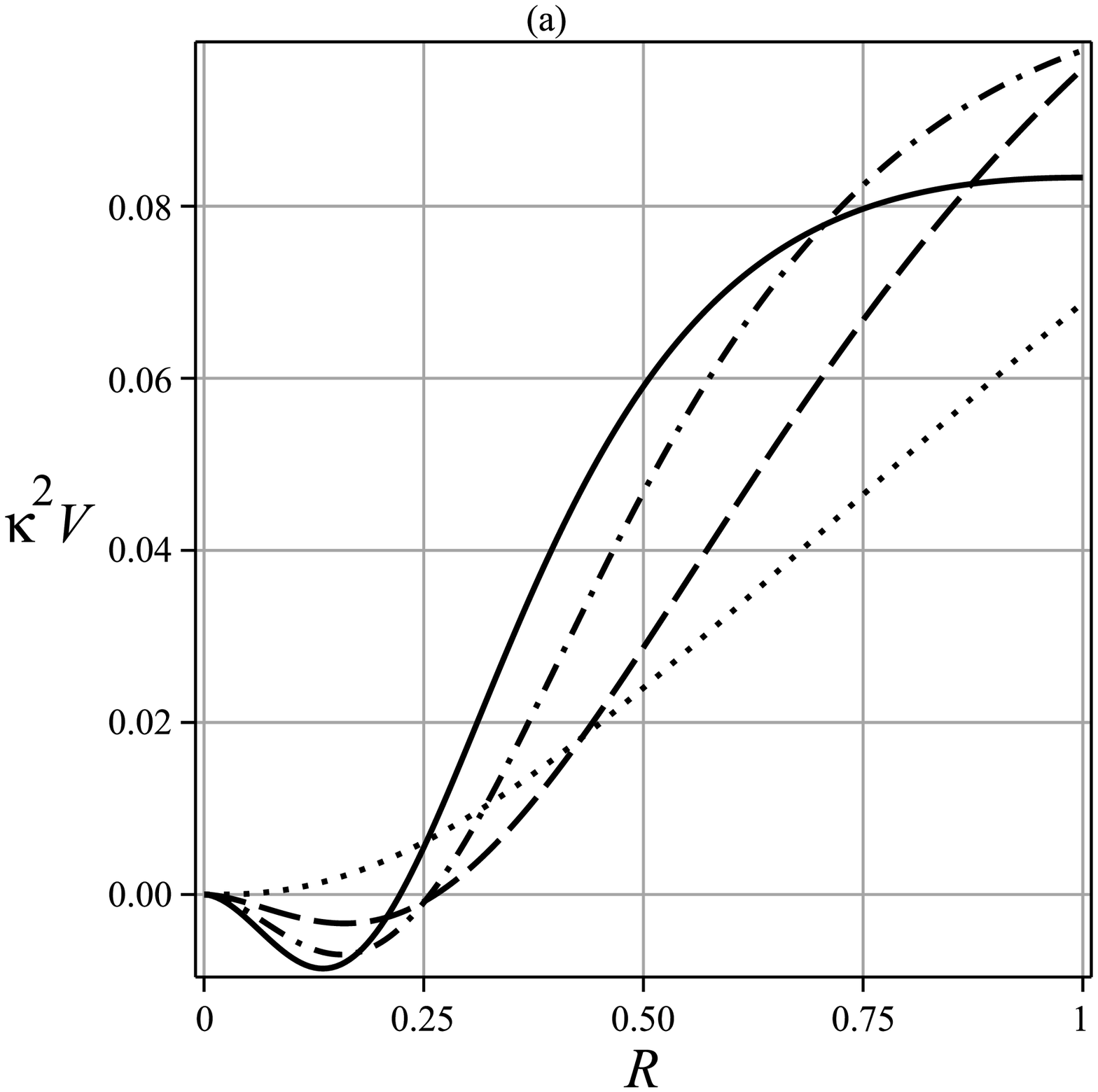}\includegraphics[width=60 mm]{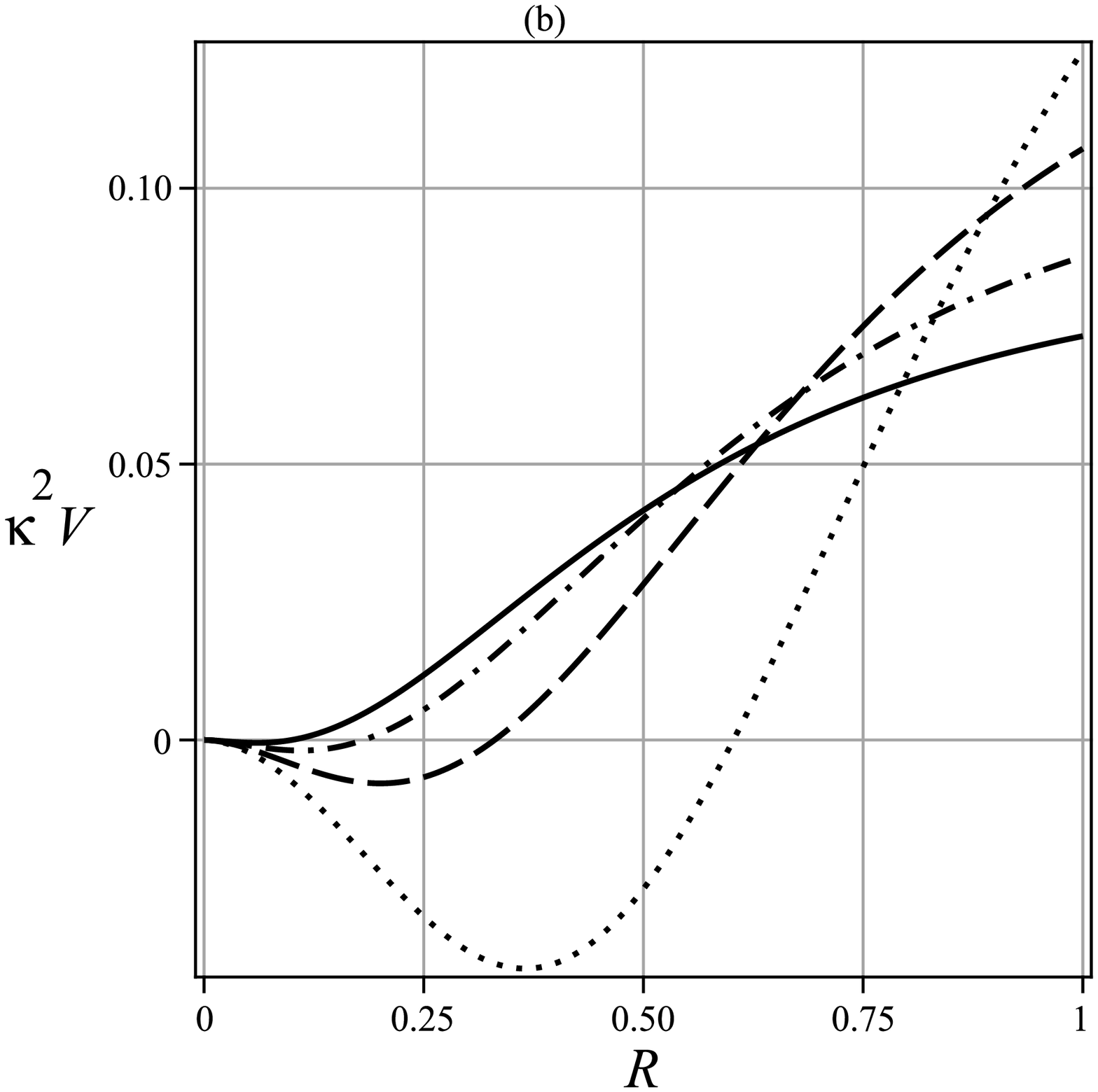}
 \end{array}$
 \end{center}
\caption{The function $\kappa^{2} V$ versus $R$. (a) $\alpha=0.5$, $\beta=0.1$ (dot), $\beta=0.4$ (dash), $\beta=0.7$ (dot dash), $\beta=1$ (solid). (b) $\beta=0.5$, $\alpha=0.1$ (dot), $\alpha=0.4$ (dash), $\alpha=0.7$ (dot dash), $\alpha=1$ (solid).}
 \label{fig2}
\end{figure}

The potential $V$ was found to be,
\begin{equation}\label{S4}
V = \frac{(\gamma-\alpha) R^2 + \beta  R^2 \ln \beta R }{2\kappa^{2} (1 + \gamma R + 2 \beta R \ln \beta
R)^{2}}.
\end{equation}
In Fig. \ref{fig2} we plotted the function $\kappa^2 V$ versus $ R $ for different values of the parameters. We can see that there is at least an extremum (minimum) obtained via $V^{\prime} =0$ which means,
\begin{equation}\label{S5}
\frac{R \left(\lambda + 2 \beta \ln (\beta
R)\right)}{\kappa^{2} \left( 1 + \gamma R  + 2 \beta R \ln (\beta
R)\right)^2} = 0,
\end{equation}

therefore,
\begin{equation}\label{S6}
2\alpha+3\beta+2\beta \ln (\beta R) = 0.
\end{equation}

Thus, using the equation (\ref{M3}) and (\ref{CCC3}) in the equation (\ref{S6})  with the condition $ \beta R < 0.5$, we found that the flat space-time is stable with $R =0$ and the curvature $R_{0} = \frac{1}{\beta}e^{-\frac{3}{2}-
\frac{\alpha}{\beta}}$ is unstable.\\
We also obtain the mass squared of a scalaron,
\begin{equation}\label{S7}
3m^{2} =\frac{1}{\lambda+ 2\beta
\ln (\beta R)} + \frac{R}{ 1 + \gamma R  + 2 \beta R \ln
(\beta R)} - \frac{4 R \left( 1 + \alpha  R + 2 \beta R
\ln (\beta R)\right)}{\left( 1 + \gamma R  + 2 \beta R \ln (\beta
R)\right)^{2}}.
\end{equation}

\begin{figure}[h!]
 \begin{center}$
 \begin{array}{cccc}
\includegraphics[width=60 mm]{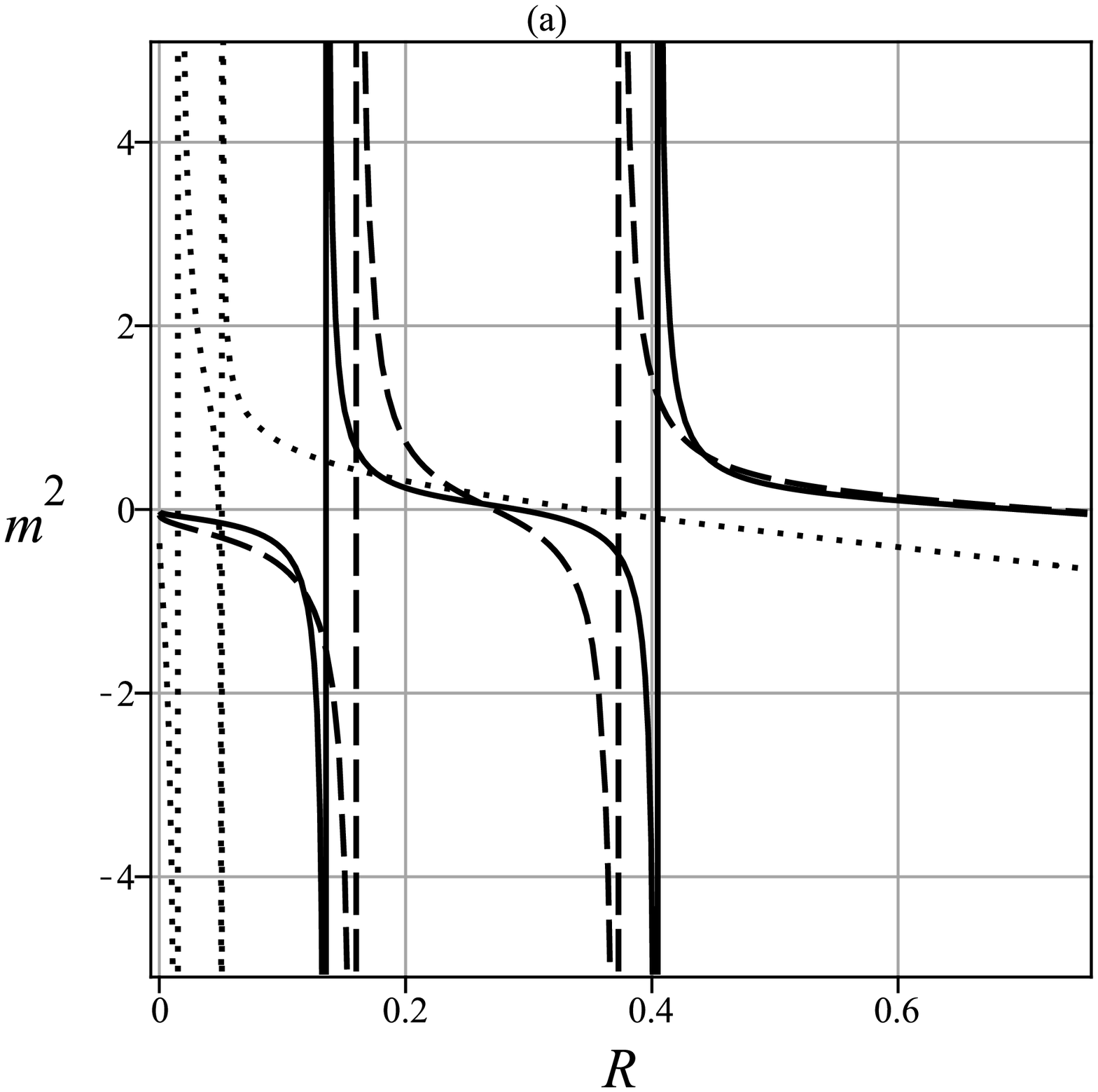}\includegraphics[width=60 mm]{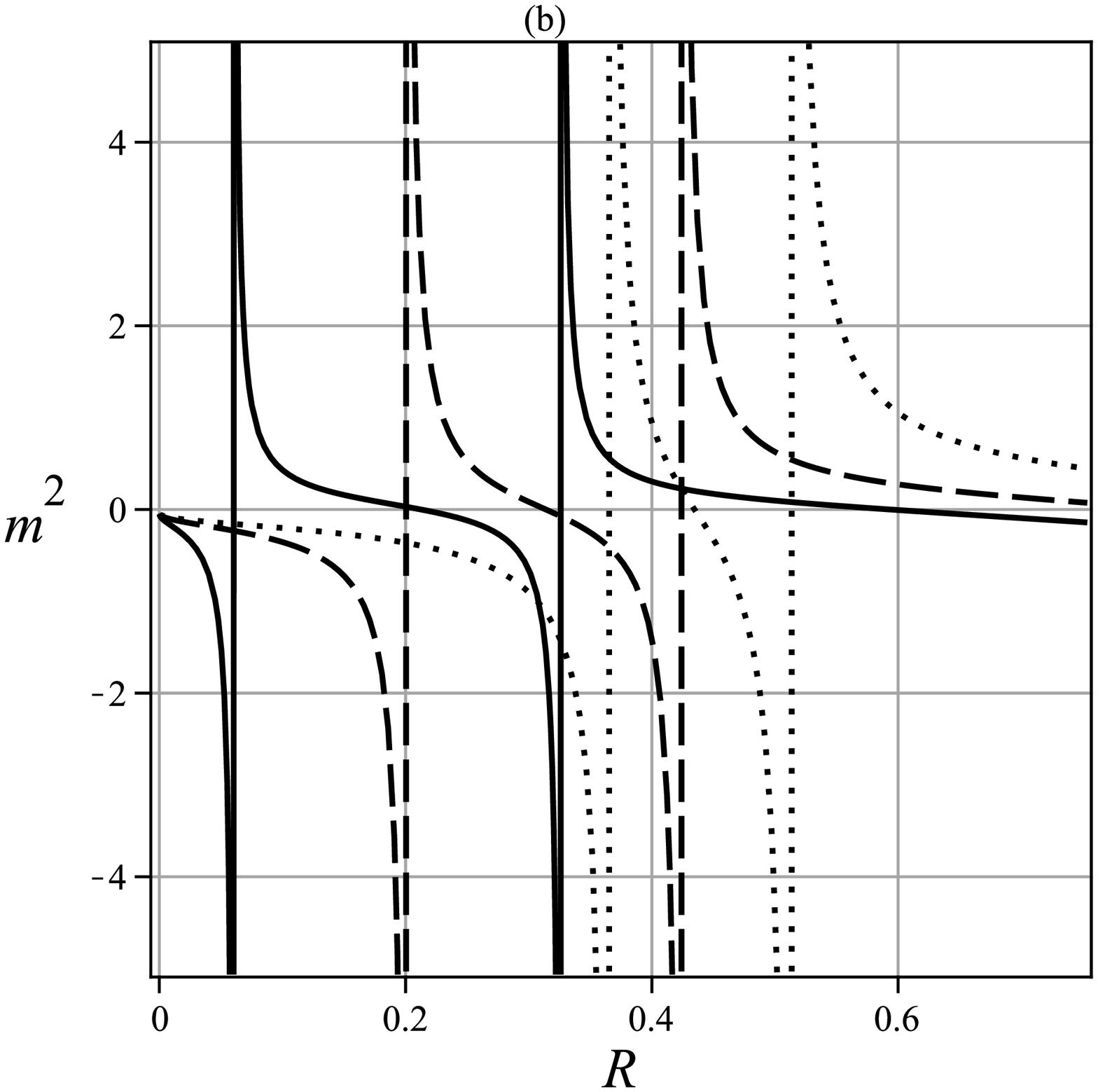}
 \end{array}$
 \end{center}
\caption{The function $m^{2} $ versus $R$. (a) $\alpha=0.5$, $\beta=0.1$ (dot), $\beta=0.4$ (dash), $\beta=1$ (solid). (b) $\beta=0.5$, $\alpha=0.1$ (dot), $\alpha=0.4$ (dash), $\alpha=1$ (solid).}
 \label{fig3}
\end{figure}

The plot of the function $ m^{2} $ versus $R$ is
given by Fig. \ref{fig3} which show periodic behavior. One can verify that $ m^{2} < 0 $ for the constant curvature
solution  $R_{0} = \frac{1}{\beta}e^{-\frac{3}{2}-
\frac{\alpha}{\beta}}$, and therefore this solution corresponds
to unstable state as it was mentioned earlier.

\section{Cosmological parameters}
We know that the corrections of $ F(R)$ gravity model are small as
compared with $GR $ for $ R \gg R_{0} $, where $R_{0} $ is a curvature
at the present time, so we have the following conditions \cite{0909.1737},
\begin{eqnarray}\label{CP1}
|F(R)-R|  \ll  R, \nonumber\\
|F^{\prime}(R)-1| \ll 1,\nonumber\\
|RF^{\prime\prime}(R)| \ll 1.
\end{eqnarray}
From the equation (\ref{I1}) we obtain,
\begin{eqnarray}\label{CP2}
\alpha R + \beta R \ln (\beta R)  \ll  1, \nonumber\\
\gamma R + 2 \beta R \ln (\beta R) \ll 1,\nonumber\\
\lambda R + 2 \beta R \ln (\beta R) \ll 1.
\end{eqnarray}

One can investigate that for $0 < \alpha < 1$ all inequalities in the equation (\ref{CP2}) are satisfied. The slow-roll parameters are given by,
\begin{eqnarray}\label{CP3}
\varepsilon & = & \frac{1}{2} M^{2}_{pl} \left (
\frac{V^{\prime}}{V}\right)^{2},\nonumber\\
\eta &=&\frac{1}{2} M^{2}_{pl}  \frac{V^{\prime \prime}}{V}.
\end{eqnarray}
For the slow-roll approximation we need the conditions $\varepsilon \ll 1$, and $ \eta \ll 1 $. One can obtain the
slow-roll parameters expressed through the curvature from the equations (\ref{S4})-(\ref{S7})
as follows,
\begin{equation}\label{CP4}
\varepsilon = \frac{1}{3} \left [\frac{\beta R -1
}{(\gamma-\alpha)R + \beta R \ln(\beta R)} \right]^{2},
\end{equation}
and
\begin{equation}\label{CP5}
\eta = \frac{2}{3} \left[\frac{1-4\beta R + \beta^2 R^2 -2
\alpha \beta R^2 - 2 \beta^2 R^2 \ln(\beta R) }{R(\lambda + 2 \beta
\ln(\beta R))(1+ \alpha R + \beta R \ln(\beta R))}\right].
\end{equation}

\begin{figure}[h!]
 \begin{center}$
 \begin{array}{cccc}
\includegraphics[width=60 mm]{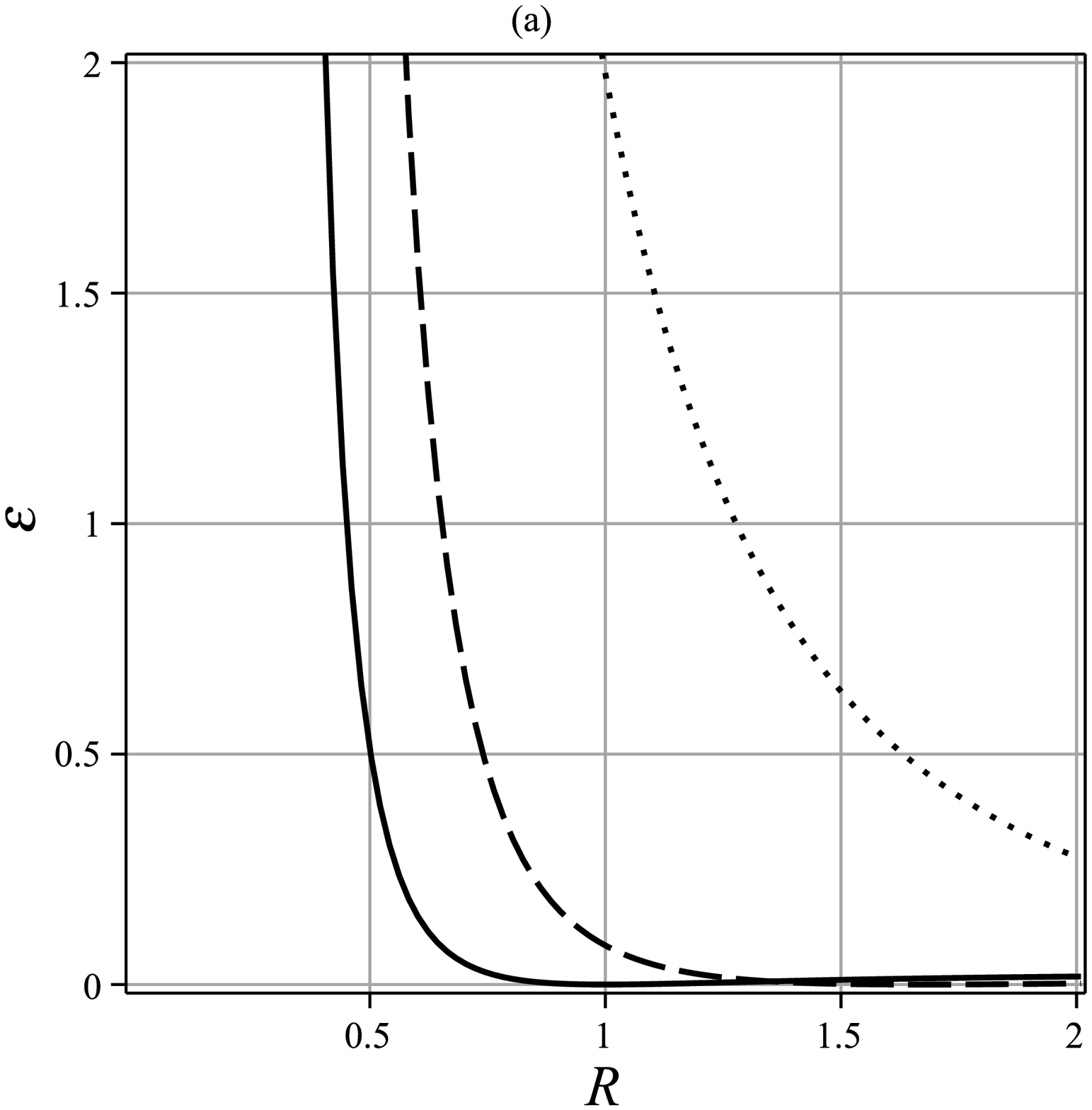}\includegraphics[width=60 mm]{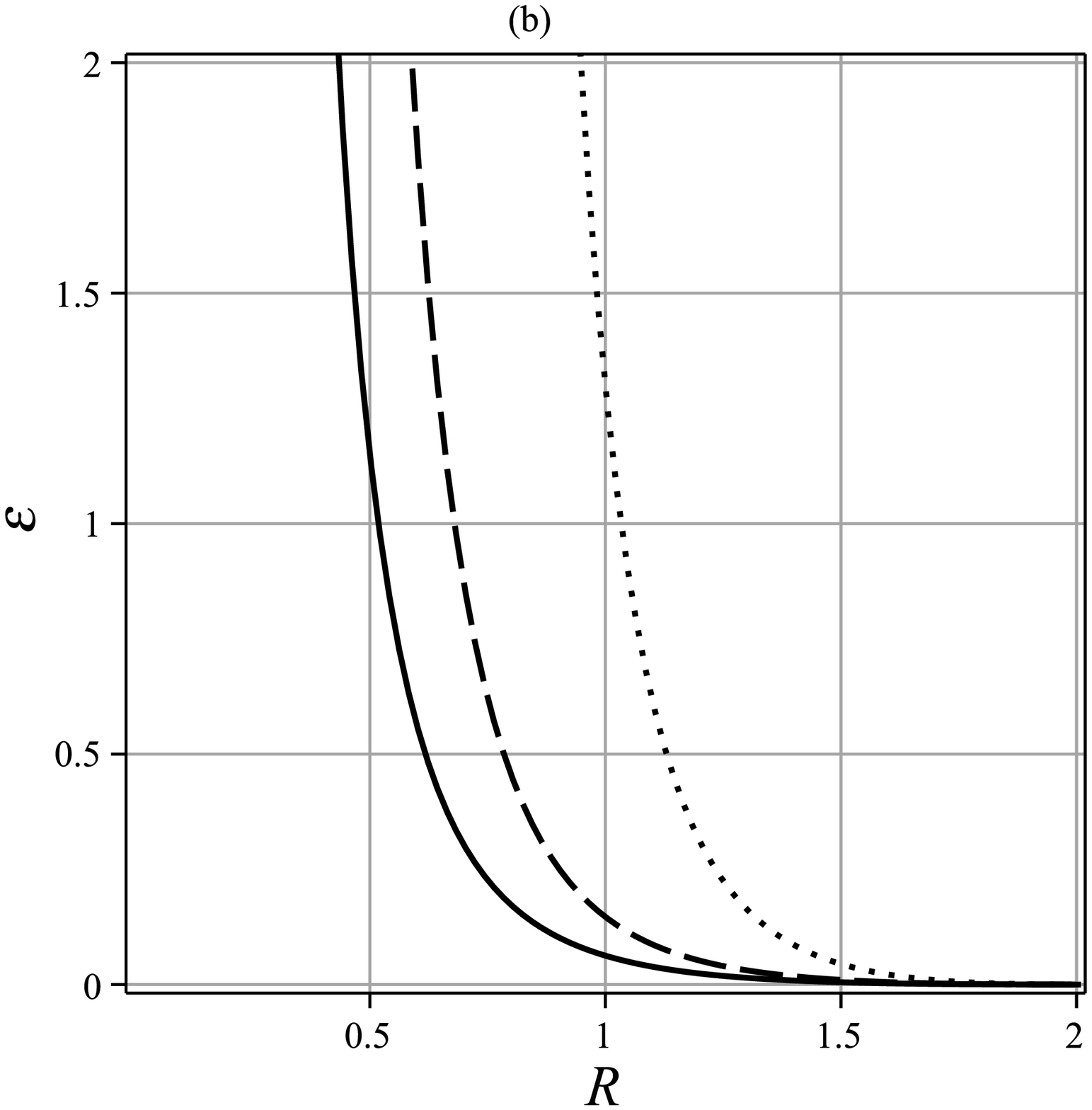}
 \end{array}$
 \end{center}
\caption{The function $\varepsilon$ versus $R$. (a) $\alpha=0.5$, $\beta=0.1$ (dot), $\beta=0.6$ (dash), $\beta=1$ (solid). (b) $\beta=0.5$, $\alpha=0.1$ (dot), $\alpha=0.6$ (dash), $\alpha=1$ (solid).}
 \label{fig4}
\end{figure}

The plot of the function $ \varepsilon $ versus $R$ is
given by Fig. \ref{fig4} for different parameters $ \beta $ and $ \alpha$.

\begin{figure}[h!]
 \begin{center}$
 \begin{array}{cccc}
\includegraphics[width=60 mm]{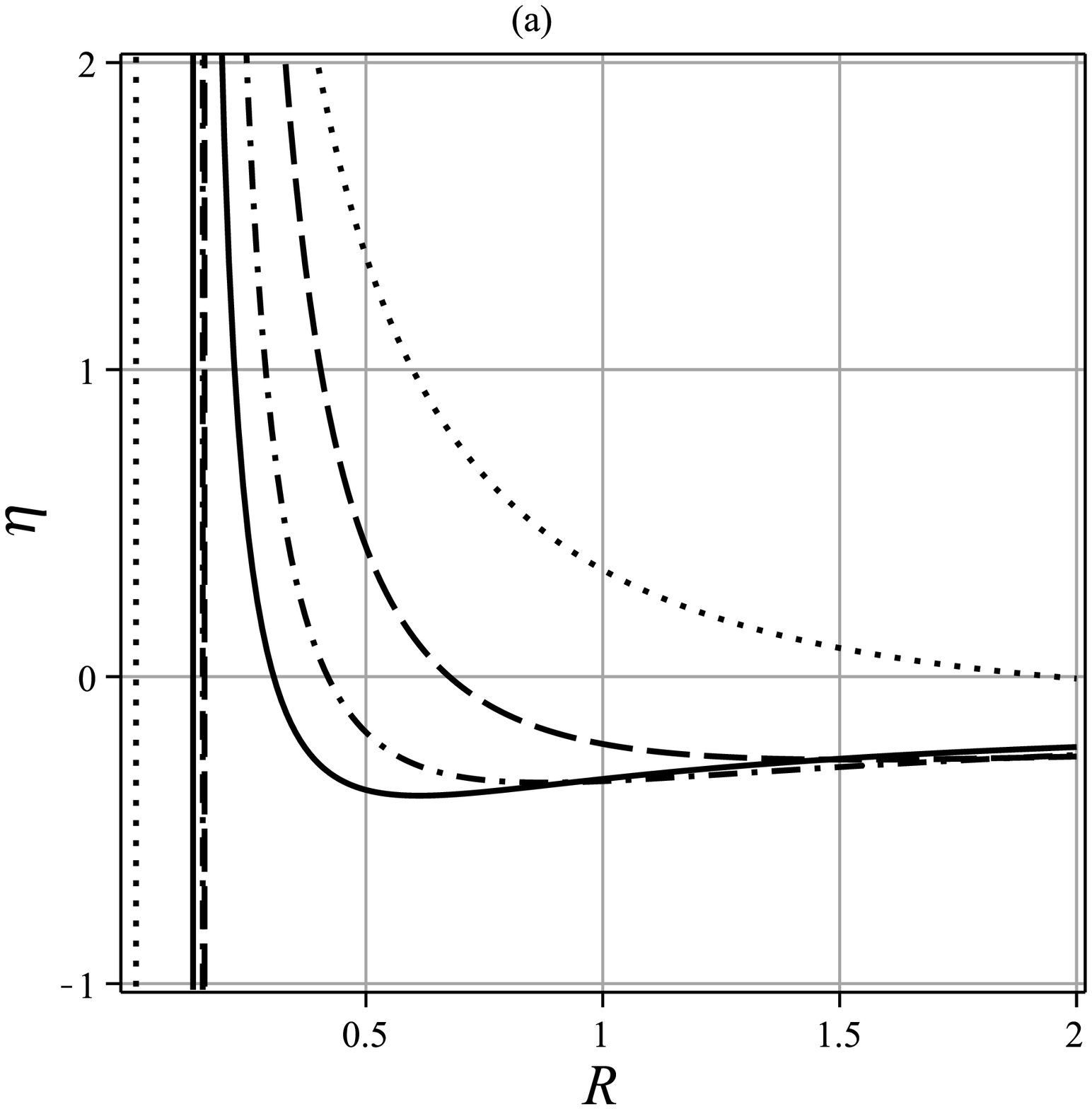}\includegraphics[width=60 mm]{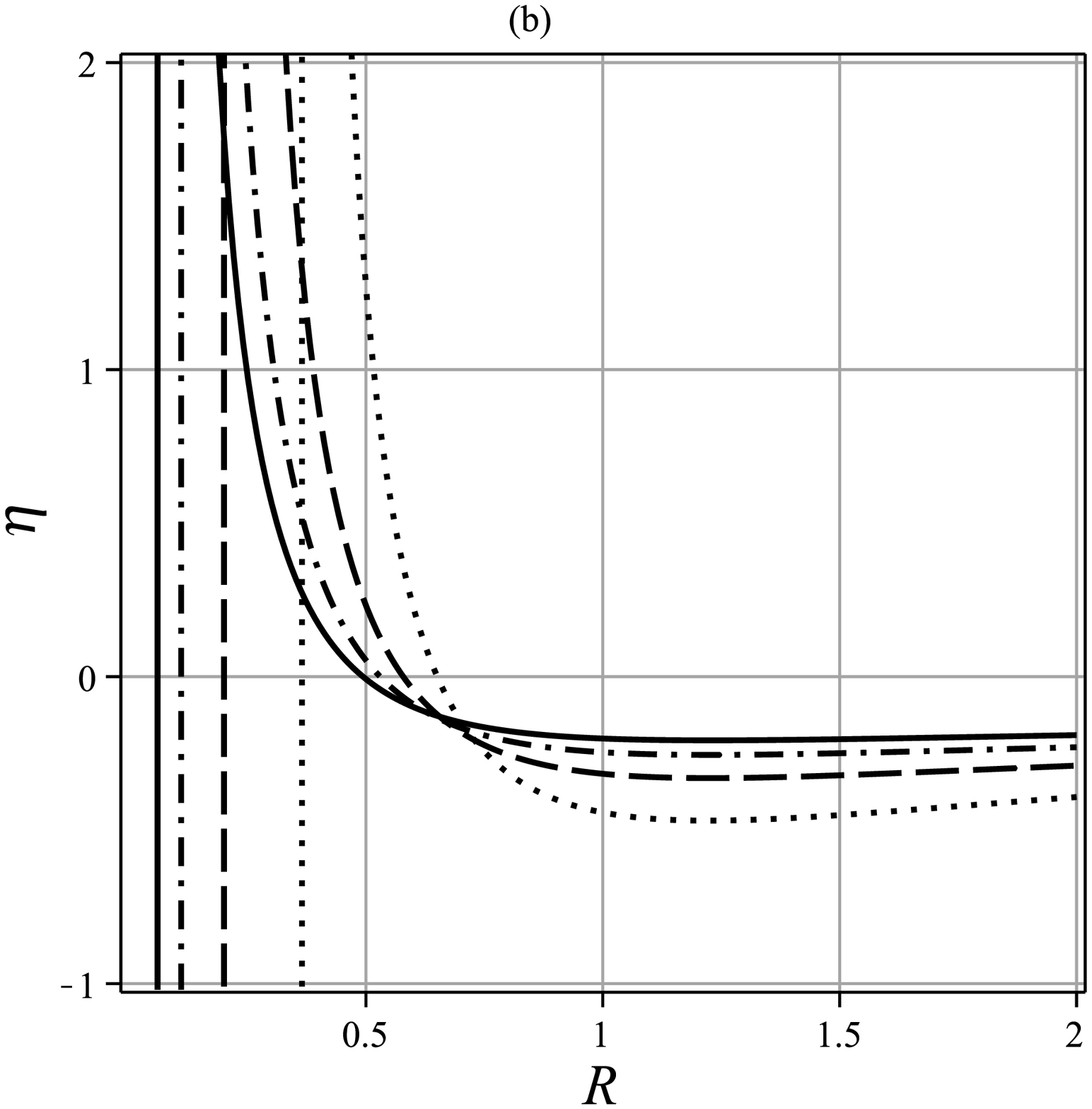}
 \end{array}$
 \end{center}
\caption{The function $\eta$ versus $R$. (a) $\alpha=0.5$, $\beta=0.1$ (dot), $\beta=0.4$ (dash), $\beta=0.7$ (dot dash), $\beta=1$ (solid). (b) $\beta=0.5$, $\alpha=0.1$ (dot), $\alpha=0.4$ (dash), $\alpha=0.7$ (dot dash), $\alpha=1$ (solid).}
 \label{fig5}
\end{figure}

The plot of the function $ \eta $ versus $R$ is given in Fig. \ref{fig5}
for different parameters $\alpha$ and $ \beta $.

The equation  $ \varepsilon = 1 $  has the solution  $ \beta R =
0.420$.  The equation $|\eta| = 1$ is satisfied when $\beta R =
0.1158$, $\beta R =  0.0142+ 0.0485 I  $ for $ 0 < \beta < 0.18 $, and when
$\beta R =  0.2429$, $ \beta R = 0.0129 - 0.7917 I$ for $ 0.18 <
\beta < 1 $. In the case of $ 0.1158 < \beta R < 1 $
and $ 0.2429 < \beta R < 1 $, we have $|\eta| < 1$.
Therefore, the slow-roll approximation, $ \varepsilon < 1$ and
$|\eta| < 1$, takes place when $0.420 < \beta R < 1$.\\

The age of the inflation can be obtained by calculating the e-fold number,
\begin{equation}\label{CP6}
N_{e}=\frac{3}{2} \int^{R_{0}} _{R_{end}}\frac{R [(\gamma -
\alpha)+ \beta \ln(\beta R)](\lambda + 2 \beta \ln(\beta R) )}{\lambda+
2\beta (1- \beta R)\ln(\beta R)- 2(\alpha + \beta)\lambda R}dRd.
\end{equation}

Here the value $ R _{end} $  corresponds to the time of the end of
inflation when $\varepsilon $ or $|\eta| $ are close to $1$. We find that selected value of $\beta R$ and $\beta$ in previous give $50<N_{e}<60$ to solve the flatness and horizon problems.\\
The index of the
scalar spectrum power law due to density perturbations is given by,
\begin{equation}\label{CP7}
n_{s} = 1 - 6 \varepsilon + 2 \eta.
\end{equation}
The tensor-to-scalar ratio is defined by,
\begin{equation}\label{CP8}
r_{s}=16 \varepsilon.
\end{equation}
The Planck experiment results \cite{2015} tell that,
\begin{eqnarray}\label{CP9}
n_{s} = 0.968 \pm 0.006, \nonumber\\
r_{s} < 0.11,
\end{eqnarray}
while adding BICEP2, Keck Array, and Planck (BKP) B-mode data yield to,
\begin{equation}\label{CP10}
r_{s} < 0.09.
\end{equation}
We can use above data to fix parameters. One can check that our selected values of $\alpha$ and $\beta$ give good result in agreement with observational data.
\section{Critical points and stability}
\begin{figure}[h!]
 \begin{center}$
 \begin{array}{cccc}
\includegraphics[width=60 mm]{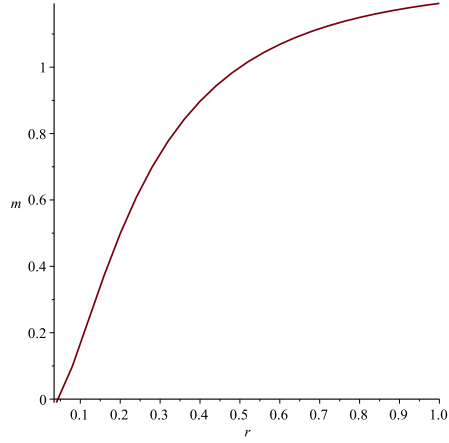}
 \end{array}$
 \end{center}
\caption{$m$ versus $r$.}
 \label{fig6}
\end{figure}

There are several ways to specify viable conditions of $F(R)$ gravity theories such as positivity of the effective gravitational coupling \cite{0907.5516}, stability of cosmological perturbations \cite{1112.4481}, equivalence principle and solar-system constraints \cite{0712.2268}, and asymptotic behavior of the $\Lambda$-CDM in the large curvature regime \cite{0705.3199}. Here, we use autonomous equations to investigate critical points and stability of the model.\\
In order to investigate critical points of equations of motion, it is useful to introduce the following dimensionless parameters \cite{0612180},

\begin{equation}\label{CPS1}
x_{1}=- \frac{\dot{F}^{\prime} (R)}{H F^{\prime}(R)}= -
\frac{(\lambda + 2 \beta \ln(\beta R))\dot{R}}{H (1+ \gamma R + 2 \beta
R \ln(\beta R))},
\end{equation}

\begin{equation}\label{CPS2}
x_{2}=- \frac{F (R)}{6 H^{2} F^{\prime}(R)}= - \frac{R+ \alpha
R^{2} + \beta R^{2} \ln(\beta R)}{6 H^{2} (1+ \gamma R + 2 \beta R
\ln(\beta R))},
\end{equation}

\begin{equation}\label{CPS3}
x_{3}=2-\frac{\dot{H}}{ H^{2}},
\end{equation}

\begin{equation}\label{CPS4}
x_{4}=- \frac{\kappa^{2}\rho_{rad}}{3 F(R) H^{2}},
\end{equation}

\begin{equation}\label{CPS5}
m =\frac{R F^{\prime \prime}  (R)}{ F^{\prime}(R)}= 1 + \frac{
 2 \beta R - 1}{1+ \gamma R + 2 \beta R \ln(\beta R)},
\end{equation}

\begin{equation}\label{CPS6}
r =- \frac{R F^{\prime }  (R)}{ F(R)}= - 2 + \frac{ 1 - \beta R
}{ 1 + \alpha R +  \beta R \ln(\beta R)},
\end{equation}

where $H$ is Hubble parameter,  and the dot denote the derivative
with respect to the time. The deceleration parameter $q$ is given by
$q = 1 - x_{3}$. The critical points for the system of equations can
be studied by the investigation of the function $m(r)$. Equations of
motion in the absence of the radiation, $\rho_{rad} = 0$ ($x_{4}=0$), with the
help of above equations can be written in the form of autonomous
equations \cite{0612180}. One can discuss the critical points of the system
of equations by the study of the function $m(r)$ which shows the
deviation from the $\Lambda$CDM model. The plot of the function
$m(r)$ is given in by Fig. \ref{fig6}.

A de Sitter point in the absence of radiation, $x_{4} = 0$, corresponds to the parameters $x_{1}= 0$, $x _{2}= -1$, and
$x_{3}= 2$ $(\dot{H} = 0, H = \frac{R}{12}, r = -2)$. This point corresponds to the constant curvature solutions. The effective equation of state (EoS)
parameter, $w_{eff}$, and the parameter of matter energy fraction,
$ \Omega _{m}$, are given for this point by the following equations,
\begin{equation}\label{CPS7}
\omega _{eff} = - 1 - \frac{2 \dot{H}}{3 H^{2}} = -1,
\end{equation}
and
\begin{equation}\label{CPS8}
\Omega _{m} = 1- x_{1} - x_{2} - x_{3} = 0,
\end{equation}

which correspond to the dark energy. This point mimics a cosmological constant
and the deceleration parameter becomes $q = -1$. The constant
curvature solution $\beta R \approx 1 $ corresponds to unstable de
Sitter space.\\

For the other critical point,
\begin{equation}\label{CPS9}
(x_{1}, x_{2},  x_{3}) = (\frac{3m}{1+m},
-\frac{1+4m}{2(1+m)^{2}},\frac{1+4m}{2(1+m)}),
\end{equation}

we can find $x _{3}= \frac{1}{2}$, $ m \approx 0$, $ r =  -1$, and EoS of a
matter era is $ \omega_{eff} = 0$ $(a = a_{0} t^{\frac{2}{3}} )$. Then,
we have a viable matter dominated epoch prior to late-time
acceleration. The equation $m = -r - 1$ has the solution $ m =
0$, $r = -1$, $ R = 0$, corresponding to this point. One can
verify using the equation (\ref{CPS6}) that $ m (r = -1) = 0$. As a result, the condition $ m (r = -1) > -1$ holds and we have
the standard matter era. Therefore, the correct description of
the standard matter era occurs in the model under consideration. The
equation $ m(r) = - r - 1 $ with the help of the equations (\ref{CPS5}) and (\ref{CPS6})
becomes
\begin{equation}\label{CPS10}
 \frac{ 1 - 2 \beta R }{1+ \gamma R + 2 \beta R \ln(\beta R)} =
 \frac{ 1 - \beta R }{ 1 + \alpha R +  \beta R \ln(\beta R)}
\end{equation}
Equation (\ref{CPS10}) possesses two solutions: the trivial solution $x =
\beta R = 0$ $(m = 0, r = - 1) $ corresponding to the second critical point discussed above, and the nontrivial solution. The nontrivial numerical
solution of the equation (\ref{CPS10}) for $ \alpha = 0.4$ gives $x \approx 3.71$, $ m
\approx 1.36$, and $ r \approx -  2.36 $.

\section{Conclusion}
In this letter, we suggested a new model of modified $F(R)$ gravity
representing the effective gravity model which can describe the
evolution of universe. Usually, the main purpose of $F(R)$ gravity models is to solve the late-time cosmic acceleration. However, it is possible to study inflation. Our main goal is to compute some inflationary parameters and compare with observational data. The constant curvature solutions, $ \beta R =0.420$, and $\beta R  = 0.919$ were obtained corresponding to the flat and the de Sitter space-time, respectively. The de
Sitter space-time gives the acceleration of universe and corresponds
to inflation. The flat space-time is stable but the de Sitter
space-time is unstable in the model and it goes with the maximum of
the effective potential. The slow-roll parameters $\varepsilon $, $
\eta $ and the e-fold number of the model were evaluated. The model
gives e-fold number $50<N _{e}<60$ characterizing the age of
inflation. Agreement of our results with observational data suggests that the logarithmic corrections are useful and may be necessary to construct successful model. We show by the analysis of critical points of autonomous
equations that the standard matter era exists and the standard
matter era conditions are satisfied. The model may be alternative to
GR, and can describe early-time inflation.\\
There are also more comprehensive model to study inflation, such as $F(R)$ proportional to polynomial inflation \cite{1309.3514} with logarithmic correction. We left this point as future work.

\end{document}